\documentclass[twocolumn]{aastex62}
\usepackage{longtable}
\usepackage{amsmath}
\usepackage{soul}
\submitjournal{ApJ}

\shorttitle{N6946-BH1 Progenitor Age \& Mass}
\shortauthors{Murphy et al.}

\newcommand{\msun}{M_{\odot}}
\newcommand{\mzams}{M_{\rm ZAMS}}
\newcommand{\bhage}{10.6^{+14.5}_{-5.9}}
\newcommand{\bhmass}{17.9^{+29.9}_{-7.6}}
\newcommand{\bhexclusion}{91\%}
\newcommand{\distmodulus}{29.47 \pm 0.079}
\newcommand{\distance}{7.83 \pm 0.29}

\begin{document}

\title{The Progenitor Age and Mass of the Black Hole Formation Candidate N6946-BH1}

\correspondingauthor{Jeremiah W.~Murphy} 
\email{jwmurphy@fsu.edu}

\author{Jeremiah W.~Murphy} 
\affiliation{Physics, Florida State University, Tallahassee, FL, USA}  

\author{Rubab Khan}
\affiliation{Astronomy, University of Washington, Seattle, WA, USA}

\author{Benjamin Williams}
\affiliation{Astronomy, University of Washington, Seattle, WA, USA}

\author{Andrew E. Dolphin} 
\affiliation{Raytheon Company, Tucson, AZ 85734, USA}

\author{Julianne Dalcanton} 
\affiliation{Astronomy, University of Washington, Seattle, WA, USA}

\author{Mariangelly D\'{i}az-Rodr\'{i}guez}
\affiliation{Physics, Florida State University, Tallahassee, FL, USA}

\begin{abstract}
The failed supernova N6946-BH1 likely formed a black hole (BH); we
age-date the surrounding population and infer an age and initial mass
for the progenitor of this BH formation candidate.  First, we use 
archival {\it Hubble Space Telescope}
imaging to extract broadband photometry of the resolved stellar populations
surrounding this event.  Using this photometry,
we fit stellar evolution models to the color-magnitude
diagrams to measure the recent star formation
history (SFH).  Modeling the photometry requires an accurate distance;
therefore, we measure the tip of the
red giant branch (TRGB) and infer a distance modulus of $\distmodulus$
to NGC~6946, or a metric distance of $\distance$ Mpc.  To estimate the
stellar population's age, we convert the SFH and uncertainties into a
probabilistic distribution for the progenitor's age.
The region in the immediate vicinity of N6946-BH1 exhibits the
youngest and most vigorous star formation for several hundred
pc. This suggests that the progenitor is not a runaway star.
From these measurements, we infer an age for the BH progenitor
of $\bhage$ Myr. Assuming that the progenitor evolved effectively as
a single star, this corresponds to an initial mass of $\bhmass$
$\msun$. Previous spectral energy distribution (SED) modeling of the progenitor suggests a mass of
$\sim$27 $\msun$.  Formally, the SED-derived mass falls within our
narrowest 68\% confidence interval; however, $\bhexclusion$ of the
probability distribtuion function
we measure lies below that mass, putting some tension between the age
and the direct-imaging results.
\end{abstract}

\section{Introduction}

In general, stellar evolution theory predicts that stars with masses above about 8 $\msun$ undergo core collapse \citep{woosley2002}.  Even though most of these
likely explode, it is not entirely clear which actually do.
Therefore, observations are crucial for identifying which
stars do or do not explode.  

Although the details vary, theories suggest that black hole (BH)
formation should be primarily seen for more massive progenitors.  At
the low-mass end,
both theory \citep{woosley2002,ugliano2012,sukhbold2016,bruenn2016,burrows2018} and observations \citep{jennings2014,smartt2015,davies2018} suggest that there is a minimum mass for
core collapse and most progenitors above this minimum mass explode and
do not form BHs \citep[7-8 $\msun$]{woosley2002}.   CCSN
simulations suggest that stars near the minimum mass explode quite
easily \citep{radice2017}, even in one-dimensional simulations.
Slightly higher mass stars, above an initial mass (the
zero-age main-sequence (MS) mass, $\mzams$) of $\sim$9 $\msun$, require the extra
boost given by multidimensional simulations
\citep{murphy08b,melson2015,roberts2016,bruenn2016,mabanta2018}.  
Therefore, according to single-star
stellar evolution theory and bolstered by the large number of
observations of low-mass progenitors of supernovae (SNe) and SN remnants
\citep{smartt2009,davies2018,maund2017,jennings2012,jennings2014,diaz-rodriguez2018}
, it is unlikely that BH formation will be seen in the lowest-mass stars.

In contrast, CCSN simulations find that the most massive stars have
difficulty exploding \citep{ugliano2012,sukhbold2016}.  Instead, a fair
fraction of these most massive stars fail to explode and form BHs.  
These investigations also suggest that the mapping between $\mzams$
and final outcome may not be entirely correlated or monotonic \citep{sukhbold2017}.
There are islands of failed SNe as a function of mass, and these
islands tend to become larger and more frequent toward the highest
masses.  These analyses present a very clear prediction: that the
distribution of progenitors that fail to explode should be heavily
skewed toward the most massive progenitors.  Testing this prediction
requires measuring a
large number of progenitor masses for CCSNe and/or BH formation
events.  

While significant progress has been made in building up statistics of
progenitor masses for successful CCSNe
\citep{smartt2009,jennings2012,jennings2014,williams2014,williams2018}, measurements of failed explosions
have only just begun. \citet{kochanek08} proposed to identify ``failed'' explosions by monitoring nearby galaxies for massive stars that
disappear without a SN and instead form a BH.  From
  this survey, \citet{gerke2015} identified several
  candidates for further monitoring.  Of these, \citet{adams2017}
identified the vanishing star, N6946-BH1, as the most
  promising BH formation candidate.  In addition,
\citet{reynolds2015} performed a similar but more limited survey
  using {\it Hubble} archival imaging of 15 galaxies.  They reported a failed SN
candidate, which will require further monitoring.
In this paper, we provide an independent
measure of the age and $\mzams$ for N6946-BH1 by age-dating the
surrounding stellar population using techniques developed in \citet{gogarten09a,murphy11a,jennings2012,jennings2014,williams2014}.

Measuring a mass for N6946-BH1 is
of the highest importance for core-collapse theory.
By modeling the
color and magnitude of the progenitor, \citet{adams2017} estimated the
progenitor's initial mass to be  $\mzams \sim 25 \msun$.  This technique of interpreting
precursor imaging is based on
using stellar evolutionary tracks to model the observed magnitude and
color of the progenitor.  If the observations include most of the
bolometric luminosity, then estimating the progenitor mass is straightforward.  For the most part, the luminosity of a post-MS star is determined by the mass of the helium core and
hydrogen-burning shell \citep{woosley2002}.  In turn, the helium core
size is set by the
zero-age MS mass.  Hence, with broad
spectral coverage, one can directly infer the progenitor mass.
However, more frequently, one only has observations in just a few spectral
bands, making modeling the luminosity extremely sensitive to mass-loss history and dust
formation in the last uncertain stages of stellar evolution.  
As a result, the mass inferred for the progenitor depends sensitively
on uncertain bolometric corrections (see, for example, the
\citet{davies2018} reanalysis of \citet{smartt2009} data on red
supergiant progenitors).

Given the importance of measuring the mass of N6946-BH1, we take
an independent approach to constraining the progenitor mass: we
age-date the stellar population, and from this age, we infer the mass of the
star that would reach the end of its lifetime at that age.
This technique has the advantage that the age is sensitive to many
phases of stellar evolution and features in the
color-magnitude diagram (CMD), not just the final uncertain stage of stellar
evolution.  For example, the tip of the MS provides an
upper limit on the age, and the presence of helium burners helps to
constrain specific ages.  The tip of the MS
is dependent upon the well-understood physics of the MS, and helium
burners are an early, more certain post-MS stage.
This technique has now been used and validated extensively \citep{gogarten09a,murphy11a,jennings2012,jennings2014,williams2014}.

In this manuscript, we age-date
the stellar population in the vicinity of N6946-BH1.  We age-date the stellar population for two reasons.  One, it
  provides an independent check on the direct technique that relies
  on modeling the uncertain physics of late-stage evolution.  Two,
  even if there is an inconsistency between the two techniques, it
  may further illuminate the evolutionary scenario -- for example, by
  pointing toward binary evolution models.  Assuming single-star
evolution for the progenitor, we then infer the progenitor mass and
compare it to the mass derived directly from the spectral energy
distribution (SED) of the progenitor.

The presentation of this manuscript is as follows.  In the method
section ( \S~\ref{sec:method}), we describe the parameters of N6946-BH1,
including those previously determined.  We also describe the method
for our age-dating technique.  In particular, we formalize how to
transform the star formation history (SFH) and uncertainties into an age distribution for the
progenitor.  In the results section (\S~\ref{sec:results}) we
determine the age, infer a progenitor mass, and consider how various
biases (such as kicks) may affect the age and mass.  
Finally, we conclude that the age is $\bhage$ Myr and the progenitor mass is
$\bhmass$ $\msun$.

\section{Method}
\label{sec:method}

To derive the progenitor mass of N6946-BH1, we employ the following
procedure.  We first derive F438W, F606W, and F814W photometry for
{\it Hubble Space Telescope} (HST) observations covering
the location of N6946-BH1 (\S~\ref{sec:method:photometry}).  We then characterize the completeness,
biases, and uncertainties in the photometry by successfully  inserting and
recovering artificial stars.  We use the resulting model of photometric
effects to derive the SFH (\S~\ref{sec:method:SFH}) by modeling the
F606W-F814W CMD for stars surrounding N6946-BH1
using the fitting program MATCH \citep{dolphin02,dolphin2012,dolphin2013}. To derive a full probability distribution for the stellar
population that produced N6946-BH1, we consider a variety of
assumptions for choice of analysis area
(\S~\ref{sec:method:photometry}), distance
(\S~\ref{sec:method:distance}), and dust (\S~\ref{sec:method:SFH}).  We then translate this age
distribution into a distribution for the progenitor mass (\S~\ref{sec:method:PDF}).

\subsection{Photometry}
\label{sec:method:photometry}

\begin{figure}[t!]
\plotone{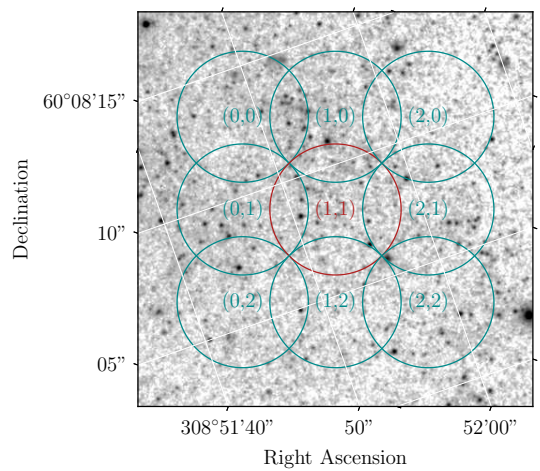}
\caption{ 
	{\it HST} F606W drizzled image showing the region surrounding the
  vanishing star and BH candidate, N6946-BH1.  
We identify all of the stars in this region,
  model the color-magnitude diagram, and infer the age of the stellar
  population surrounding the vanisher.  The central red circle is a
  region surrounding the vanisher with a radius of 100 pc (the distance to
  NGC~6496 is $\distance$ Mpc). It is possible that the vanisher was born in
  another nearby region and received a kick to its current location.
  Later, we model this uncertainty by inferring the star formation
  region for the eight surrounding regions (blue circles), each with a radius of 100 pc.}
\label{fig:FindingChart}
\end{figure}

We used pipeline-calibrated HST Wide Field Camera 3
(WFC3) UVIS images from several general observing campaigns.  See
Table~\ref{fig:SFHdatasets} for the proposal IDs, PI, filters, and total exposure
times. 
Figure~\ref{fig:FindingChart} shows an image for the region
surrounding N6946-BH1.  The location of the progenitor for
N6946-BH1 was at the center of the red circle, R.A. 20:35:27.56 and decl. +60:08:08.29.  The GO-13392 images were taken on
2014 February 21, the GO-14266 images were observed on 2015 October 8, and
the GO-14786 images were taken on 2016 October 26, which is 5-7 yr after the observations in which the N6946-BH1 had
disappeared (2009).  

We derived photometry from the images using a modified version of the
PHAT pipeline, which is built around DOLPHOT \citep{dolphin2002,williams2014b}.  We selected
stars with signal-to-noise ratio ${\rm SNR} > 4$, sharpness squared~$<0.15$, and
crowding~$<1.3$.  
We combined the photometry from all epochs of imaging; the resulting data reach a
depth of 27.64, 28.23, and 26.74 in F438W, F606W, and F814W, respectively.  These
correspond to the brightness of a solar-metallicity MS star
with mass 12, 9, and 15 $\msun$ respectively \citep{marigo2017}. To
convert these magnitude depths into MS mass, we assume a distance to NGC~6946 of $\distance$ Mpc
(see the derivation of the distance in \S~\ref{sec:method:distance}) and total foreground extinction of $A(V) = 0.94$ \citep{schlafly2011,adams2017}.

\begin{table}[h!]
\caption{Data sets for inferring the SFH in the
  vicinity of the vanisher, N6946-BH1}
\begin{tabular}{cccc}
\hline
Prop ID & PI & Filter & Exposure Time (s) \\
\hline
13392  & B. Sugerman  & F438W  &   2768 \\
13392  & B. Sugerman  & F606W   &  1744 \\
13392  & B. Sugerman  & F814W   &  1744 \\
14266  & C. Kochanek  & F606W   &  1233 \\
14266  & C. Kochanek  & F814W   &  1233 \\
14786  & B. Williams  & F438W   &  2670 \\
14786  & B. Williams  & F606W   &  2800 \\
\hline
\end{tabular}
\tablecomments{The first column gives the
  proposal ID, the second column provides the PI, the third column is
  the WFC3/UVIS filter, and the fourth column is the total exposure
  time.}
\label{fig:SFHdatasets}
\end{table}

\citet{gerke2015} measured the photometry for the progenitor of
N6946-BH1 from HST archival images.  The resulting magnitudes are
$23.09 \pm 0.01$ in F606W and $20.77 \pm 0.01$ in F814W.

Figure~\ref{fig:FindingChart} also shows regions for
which we characterize the stellar populations.  The red circle is centered on the vanisher.  We adopt a radius of 100 pc for these
regions.  Early experiments on the size of the region \citep{gogarten09a} suggested
that $\sim$50 pc is big enough to include a significant number of MS stars for age-dating but not too big to include too many
stars from neighboring populations.  However, that exploration was for
one region of one galaxy, and the optimum size will likely differ
depending upon the local stellar density.  For now, we adopt 100 pc
radii and leave a more thorough analysis on the dependence of size
for future work.  In addition to the region centered on the vanisher,
we also infer the age of eight nearby 100 pc radius regions
(blue circles), which allows us to consider scenarios in which the
vanisher was ejected from a neighboring star-forming region.  For a
characteristic ejected velocity of $\sim$10 km s$^{-1}$ and a lifetime of
$\sim$10 Myr, the vanisher could have traveled 100 pc before
undergoing core collapse.

\begin{figure}[t!]
\plotone{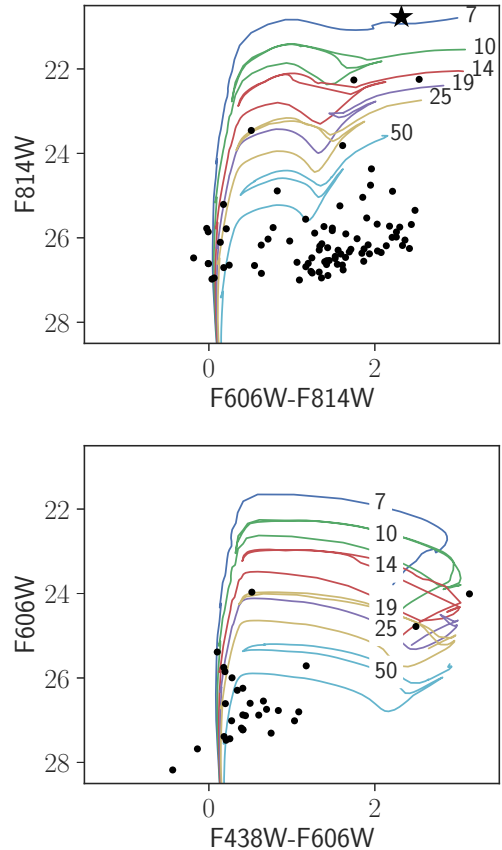}
\caption{The CMDs for a region within 100 pc of the
vanisher.  The bottom panel shows F606W vs. F438W-F606W, and the top
panel shows F814W vs. F606W-F814W.  The star in the top panel
  shows the photometry for N6946-BH1 before it vanished \citep{gerke2015}.  For comparison, both plots also show
model isochrones for 7, 10, 14, 19, 25, and 50 Myr. The distance
is $\distance$ Mpc with an $A(V)$ of 0.94.  We use MATCH to infer the
SFH from the F606W-F814W CMD.}
\label{fig:cmds}
\end{figure}

In Figure~\ref{fig:cmds}, we plot  the F814W versus F606W-F814W (top panel)
and F606W versus F438W-F606W (bottom panel) CMDs for the central region colocated
with the vanisher.  The star in the top plot represents the photometry
of N6946-BH1 before it vanished in the optical bands
\citep{gerke2015}.  Before N6946-BH1 vanished, it was the brightest
star in the F814W filter in this region.  In
fact, it is the brightest star within 300 pc.  Later, we find
  the SFH both with and without the progenitor; it
  made no real difference in the resulting SFH.  For reference, we plot isochrones for six ages
(7, 10, 14, 19, 25, and 50 Myr) assuming solar metallicity and a
foreground reddening of $A(V) = 0.94$ and $R(V) = 3.1$
\citep{marigo2017}.  

We infer the SFH for both F438W versus F438W-F606W and F606W
  versus F606W-F814W CMDs independently, but we only report the more
  accurate SFH for F606W vs. F606W-F814W.  In principle, one may use MATCH
  to infer an SFH based upon all three bands.  However, MATCH
  is not able to calculate the uncertainty in the SFH based upon three
  bands.  It calculates the uncertainty for two or four bands.  We
  find that the SFH is a little more accurate when using the F606W and F814W bands.
Furthermore, the progenitor is extremely red and only has
upper limits for F438W \citep{adams2017}.  To give a broader
  sense of the CMD, we report both CMDs, but for inferring SFHs and
  ages, we primarily report ages using the F606W and F814W CMD.

\subsection{Distance to NGC 6946}
\label{sec:method:distance}

When inferring the SFH, one must convert model
luminosities into model fluxes; this requires an accurate distance to
NGC~6946.  NGC~6946 has had roughly 10 observed SNe in the last 50
yr, so deriving an accurate distance to this one galaxy will
greatly improve the statistics of many local SNe.  Here we describe
our method to calculate the distance to NGC~6946 using the tip
of the red giant branch (TRGB).

There are 32 distance estimates in the NASA/IPAC Extragalactic
Database (NED) with a mean of 5.5 Mpc and a standard deviation of 1.5 Mpc.  A
variety of techniques were used to determine these distances; they include the brightest blue stars, Tully-Fischer, TRGB,
and the expanding photosphere for SN Type II methods.  The reported uncertainties on
most of the distance moduli are of order 0.1.  
\citet{gerke2015} and \citet{adams2017}  used a distance of 5.9~Mpc derived from the
brightest blue stars method \citep{karachentsev2000}.  However, we note that
\citet{karachentsev2000} actually reported a distance of 6.8 Mpc for
NGC~6946. The confusion in distance from
\citet{karachentsev2000} is understandable.  The text of that
manuscript only refers to the average distance to the NGC~6946 group
(5.9 Mpc),
but Table 4 of \citet{karachentsev2000} explicitly gives the distance to NGC~6946 as 6.8 Mpc.
More recently, \citet{tikhonov2014} used the
expanding photosphere method and inferred a distance of $6.72 \pm 0.15$
Mpc.  In any case, these
distance measures are fairly uncertain compared to a well-calibrated
TRGB distance estimate.

\citet{mcquinn2016} described a precise method for determining the
distance from the TRGB.  Our first step is to use HST images to calculate accurate photometry for a large number of stars
in NGC~6946.  The images were taken in the F606W and F814W bands by
program HST-GO-14786 (PI: B. Williams).  They contain 5470 s of
exposure in F606W and 5538 s of exposure in F814W.  They reach 27.5
in F814W and 28.6 in F606W.  They were observed on 2016-11-02 pointed
at R.A. = 308.8530225, decl. = 60.040649634.
We derived F606W and F814W photometry from the images using a modified version of the
PHAT pipeline, which is built around DOLPHOT
\citep{dolphin2002,williams2014b}.  The TRGB location was measured using
a Bayesian maximum-likelihood technique described in
\citet{mcquinn2016}, which is based on \citet{makarov2006}.

The measured TRGB magnitude is $26.004 \pm 0.035$ at a color of
${\rm F606W}-{\rm F814W}=1.71$.  The fit is clean; in particular, there are no
secondary peaks that might indicate issues.  With a total foreground
extinction of $A(V) = 0.94$ \citep{schlafly2011,adams2017}, the
corresponding reddenings are $A({\rm F606W}) = 0.88$ and $A({\rm F814W}) =
0.57$.  Therefore, the reddened-corrected TRGB is 25.44 with a color
of 1.40.  \citet{rizzi2007} provided a calibration for the absolute
magnitude for the TRGB: 
$M({\rm ACS}\, {\rm F814W}) = -4.06 + 0.20[({\rm F606W}-{\rm F814W}) -1.23]$.
Given the reddening-corrected color, the F814W absolute
magnitude is $M({\rm ACS}\, {\rm F814W}) = - 4.03.$
Sources of uncertainty in the \citet{rizzi2007} calibration and its
underlying horizontal-branch (HB) calibration \citep{carretta2000}
add 0.05 mag of uncertainty, while an adopted 10\% uncertainty
in the foreground extinction adds another 0.05 mag of
uncertainty \citep{schlafly2011}.  Combined with the measurement uncertainty, we calculate
a distance modulus of $m-M = \distmodulus$.  The corresponding metric
distance is $\distance$ Mpc. 

This new distance for NGC~6946 suggests that the vanisher was
intrinsically 1.3 times more luminous than assumed in
\citet{adams2017}.  According to the \citet{marigo2017} models, the larger distance would shift their mass
estimate from $\sim$25 $\msun$ to $\sim$27 $\msun$.  As a
  rough measure of systematics in the stellar evolution
  models, \citet{adams2017} also estimated the mass using rotating
  models.  Their SED-derived mass from the rotating models was
  $\sim$22 $\msun$.  With the new distance, this would shift to
  $\sim$24 $\msun$.

\subsection{Star Formation History}
\label{sec:method:SFH}

We derive SFHs from the F606W-F814W CMD using the program MATCH \citep{dolphin02,dolphin2012,dolphin2013},
which generates model CMDs that include the effects of observational
errors (as characterizing the artificial star
tests), foreground and internal dust extinction, and distance and then
adjusts the SFHs to maximize the likelihood of the observed CMD.  We
run MATCH using logarithmically spaced age bins, $\Delta \log_{10}(t \text{yr}^{-1}) = 0.05$, and the
left edge of the minimum age bin is $\log_{10}(t \text{yr}^{-1}) = 6.6$).
Using the TRGB distance, we fix the
distance to be 7.83 Mpc.



We assume an extinction law with $R(V) = A(V)/E(B-V) = 3.1$
\citep{cardelli1989,odonnell1994} and use the same galactic foreground
reddening as \citet{adams2017} of $E(B-V) = 0.303$ \citep{schlafly2011} to fix the foreground extinction in MATCH to be $A(V) = 0.94$.  However, we allow MATCH to fit for
a distribution of internal extinctions; the default model in MATCH
uses a top-hat distribution
of extinction, with width $dA_V$.  This is designed to approximate the
behavior of young stars enshrouded in a layer of patchy dust.  We
infer $dA_V$ for each region in Figure~\ref{fig:FindingChart}.  Starting from
the top left, the values for $dA_V$ are 0.00, 0.19, and 0.00 (top row);
0.70, 0.39, and 0.00 (middle row); and 0.05, 0.17, and 0.00 (bottom
row).  Taking the internal extinction as a tracer of the presence of
molecular gas, one might expect the youngest population to be in the
region containing the vanisher and the region to its left in
Figure~\ref{fig:FindingChart}. 

\subsection{Deriving the Age PDF from the SFH and Uncertainties}
\label{sec:method:PDF}

The SFH provides the stellar mass formed as a function of age.  To
estimate the age probability distribution function (PDF), we assume
that the probability is proportional to the mass of stars formed.  If
one knows the exact SFH, then the age PDF would simply be proportional
to the best-fit SFH.  However, MATCH provides a hybrid Markov chain
Monte Carlo (MCMC) algorithm, which
returns a distribution of SFHs given the photometric data of the
surrounding stars.  Therefore, one must derive the age distribution
given the distribution of accepted SFHs.

The goal is to derive the marginalized age distribution $P(\tau)$,
given the distribution of SFHs ($\{S_j(\tau)\}$) that MATCH returns;
$j$ indexes each MCMC sample.
When converting an SFH to a PDF, one must choose a maximum age for the normalization.  A reasonable choice for this maximum
age is the maximum age for core collapse, $T_{\rm max}$.  Given these
assumptions and dependencies, the joint probability for observing a
burst of star formation is $P(\tau,S,T_{\rm max})$.  Marginalizing this
distribution gives the age PDF
\begin{equation}
\label{eq:marginalizedpdf1}
P(\tau) = \int P(\tau,S,T_{\rm max}) \, dS \, dT_{\rm max} \, .
\end{equation}

To derive the joint probability distribution, we use the conditional
probability theorem, which states that 
\begin{equation}
\label{eq:jointdistribution}
P(\tau,S,T_{\rm max}) = P(\tau | S, T_{\rm max}) P(S) P(T_{\rm max})
\, .
\end{equation}
Here $P(\tau | S, T_{\rm max})$ is the likelihood for $\tau$ given the SFH,
$S$, and maximum age in converting the SFH into a PDF, $T_{\rm max}$:
\begin{equation}
P(\tau | S, T_{\rm max}) = \frac{S(\tau)}{M_{\star}(T_{\rm max})} \, ,
\end{equation}
where $M_{\star} = \int_0^{T_{\rm max}} S(\tau) \, d\tau$ is the total
stellar mass formed in the last $T_{\rm max}$ years.
Here $P(S)$ represents the distribution of SFHs from MATCH hybrid MCMC
runs, and $P(T_{\rm max})$ represents a prior on $T_{\rm max}$.
\citet{diaz-rodriguez2018} recently found the maximum age for core
collapse to be $50.3^{+2.5}_{-0.5}$ Myr, and one could use this
distribution as the prior.  However, this distribution is preliminary
in that the \citet{diaz-rodriguez2018} analysis does not include the
uncertainties in the SFH in the conversion from SFH to PDF.  The work
of this section will provide the foundation for a more thorough
analysis.  Therefore, for the purposes of this manuscript, we will
employ the same prior that was used in previous studies
\citep{gogarten09a,murphy11a,williams2014}; we will only
consider ages below $\widetilde{T}_{\rm max}$ in the conversion from an SFH
to a PDF.  Hence, the prior for $T_{\rm max}$ is a delta function at
$\widetilde{T}_{\rm max}$: 
\begin{equation}
\label{eq:tmaxprior}
P(T_{\rm max}) = \delta(T_{\rm max} - \widetilde{T}_{\rm max}) \, ,
\end{equation}
where $\widetilde{T}_{\rm max} = 50$ Myr.  This prior simplifies the
marginalization in eq.~(\ref{eq:marginalizedpdf1}) to
\begin{equation}
\label{eq:marginalizedpdf2}
P(\tau) = \int P(\tau|S,\widetilde{T}_{\rm max})P(S) \, dS \, .
\end{equation}

MATCH returns discreet samples of SFHs with star formation rates
(SFRs) at discreet ages;
hence, the age PDF will be discreet.
In general, samples returned by MCMC algorithms are designed so that the density
of the samples is proportional to the probability density function.
Therefore, the probability of each SFH sample is
\begin{equation}
P(S) = \frac{1}{N} \, ,
\end{equation}
where $N$ is the number of samples.  The discreet
marginalized PDF for the age then becomes
\begin{equation}
\label{eq:discreetpdf}
P(\tau_i) = \frac{1}{N}\sum_{j=1}^{N}
\frac{S_j(\tau_i)}{M_j(\widetilde{T}_{\rm max})} \, .
\end{equation}
The index $i$ specifies the age bin, $j$ specifies the MCMC sample,
and $M_j(\widetilde{T}_{\rm max})$ is the total stellar mass formed in the last
$\widetilde{T}_{\rm max}$ yr.  If the total mass formed is the same for
each sample, then the discreet marginalized PDF is equal to the
average SFH divided by the total stellar mass formed in the last
$\widetilde{T}_{\rm max}$ years.

\subsection{Reporting the Confidence Interval}

In this manuscript, we report the most likely age and the narrowest
68\% confidence interval (CI).  Alternatively, one might report the
68\% CI bounded by the 16th and 84th percentiles.  However, this
often-used technique is prone to several biases when applied to real
data.  If the PDF has exceptionally large wings, as real data often
do, then the 16th and 84th percentiles tend to be centered around
the median, not the mode.
The age PDF in Figure~\ref{fig:pdf_sfh} has a relatively narrow mode
but very large wings.  Therefore, if we simply report the 16th and
84th percentiles, then the CI will be strongly biased to the middle of
the age range that we consider.  Effectively, this CI will be centered
around $\widetilde{T}_{\rm max}/2$, making the CI strongly dependent
upon the maximum age chosen for normalization.  Reporting the narrowest 68\% CI mitigates these problems.  The narrowest 68\% CI is
less sensitive to the large wings in the PDF, is much less dependent on
$\widetilde{T}_{\rm max}$, and naturally centers on the mode of the distribution.

\section{Results and Discussion}
\label{sec:results}

\begin{figure}[t!]
\epsscale{1.2}
\plotone{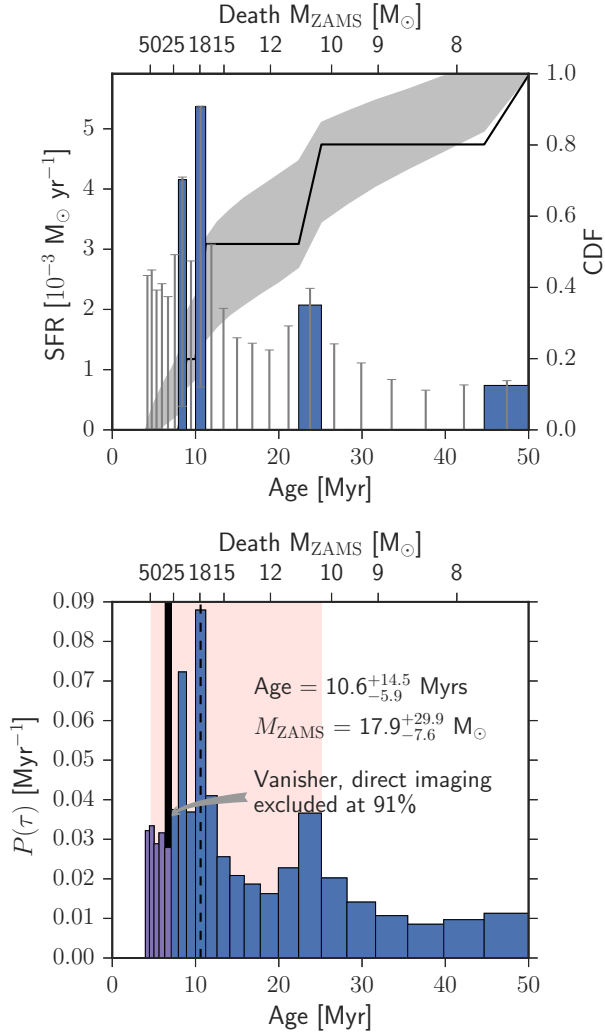}
\caption{Age distribution for the 100 pc region surrounding the
  vanisher, N6946-BH1.  The top panel shows the SFH. The blue bars show the best-fit SFH from MATCH, and
  the gray uncertainties represent the distribution of SFHs from MCMC
  runs.  The top panel also shows the CDF in the last 50 Myr (black line with 68\% CI in
  gray).  The bottom panel shows the marginalized age distribution, the
  most likely age (vertical dashed line), and the narrowest 68\%
  CI (rose band).  We
  find the most likely age to be $\bhage$ Myr, and the corresponding birth
  mass is $\mzams = \bhmass$ $\msun$.  The vertical black bar shows
  the SED-derived mass \citep{adams2017}.  The
	SED-derived mass is different than the best-fit age-derived mass,
	but the results are formally consistent.}
\label{fig:pdf_sfh}
\epsscale{1.0}
\end{figure}

Figure~\ref{fig:pdf_sfh} shows the SFH and the age distribution,
$P(\tau)$, based upon the F606W-F814W CMD for the central 100
pc radius region in
the vicinity of N6946-BH1.  The top panel shows the best-fit SFH
(blue bars) and uncertainties (gray regions).  The uncertainties represent the distribution of
MCMC sample SFHs.  The top panel also shows the best-fit cumulative
distribution function (CDF) within the last 50 Myr (black line).  The gray
region represents the 68\% CI on the cumulative distribution.
Along the top horizontal axis, this figure shows the $\mzams$ that
dies at the age of the lower horizontal axis.  For this age-to-mass
mapping, we use the results of PARSEC v1.2S isochrones \citep{marigo2017}. 
There are two bursts of star formation within the last 50 Myr; the
highest SFR is found in the most recent one, at $\sim$10 Myr.  This age
corresponds to a progenitor mass of $\sim$18 $\msun$.  The next oldest burst,
at $\sim$23 Myr, would correspond to a much lower mass ($\sim$10
$\msun$).

To test the sensitivity of the SFH to the presence of the vanisher, we
derived the SFH for two scenarios.  In one, we included the
photometric properties of the vanisher.  In the second, we omitted the
vanisher.  The resulting SFHs and uncertainties were nearly
identical.  These results suggest that the SFH is not sensitive to the
presence or absence of the progenitor.

The bottom panel of Figure~\ref{fig:pdf_sfh} shows the marginalized age distribution for
the stellar population, $P(\tau)$ (blue bars).  Assuming that the
progenitor evolved as a single star and was born with the surrounding
population, then the most likely age for the progenitor is $\bhage$ Myr.
This corresponds to an $\mzams = \bhmass$.  The dashed line shows the
most likely age and mass, and the rose-colored region shows the
narrowest 68\% CI.  The vertical black bar shows the
SED-derived mass estimate for the vanisher \citep{adams2017}.
Formally, the SED-derived mass falls within the narrowest 68\% CI; however, $\bhexclusion$ of the PDF we measure
lies below that mass, putting some tension between the age and the
direct-imaging results.

The narrowest 68\% CI for the age lies
  between 4.7 and 25.1 Myr.  In contrast, the 16\% and 84\%
  percentiles are 8.4 and 34.1 Myr.  While the SED-derived mass and
  age lie within the narrowest 68\% CI, it lies outside the 68\% range
  based upon the percentiles.  The narrowest 68\% CI is a more
  reliable estimator of the distribution's width near the mode because the
  large tails heavily bias the calculation of
  the percentiles toward the middle of the distribution.

\begin{figure}[t!]
\epsscale{1.2}
\plotone{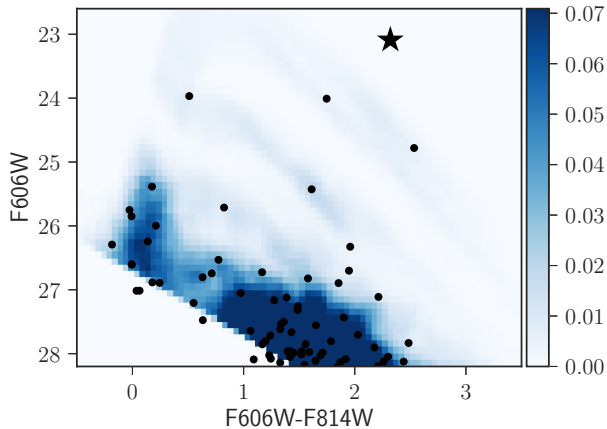}
\caption{Comparison between the Hess diagram (modeled CMD) and the
  observed CMD.  The points show the observed CMD and the F606W magnitude
  vs. the color, F606W-F814W.   The color map represents the model.
  The blue shading presents the
  expected number of stars per Hess bin; each Hess bin has a width in
  color of 0.05 and a height in magnitude of 0.1.   This modeled CMD
  represents the best-fit SFH in the top panel of
  Figure~\ref{fig:pdf_sfh}.  The model accurately represents the observed MS, including the MS turnoff, the
  red helium burners (the four or five red and yellow supergiants),
  and the swarm of red giants (lower right) associated with older
  ages.  The star represents the magnitude and color of the vanisher.}
\label{fig:cmd_model}
\epsscale{1.0}
\end{figure}

Figure~\ref{fig:cmd_model} compares the model CMD (Hess diagram) with
the observed F606W versus F606W-F814W CMD.  The background blue-scale image
represents the Hess diagram for the best-fit model in the top panel of Figure~\ref{fig:pdf_sfh}.  The
shade of blue represents the number of expected stars per Hess bin, where each Hess
bin is 0.1 mag in height and 0.05 mag in the width
(color).  The black circles are the observed stars within 100 pc of the
vanisher, and the star shows the color and magnitude of the vanisher.
In general, the model accurately represents the density of stars along
the MS and the helium burners.  The low-luminosity red giants in the
lower right contribute to ages in the star formation that are older
than 50 Myr.  The upper MS and the few bright yellow and red supergiants constrain the young ages.

It is possible that the progenitor is not
coeval with the surrounding stellar population, which could impact
the certainty of the age and progenitor mass.  For example, stellar
dispersion and kicks are two mechanisms that may cause the progenitor
to no longer be located with its coeval birth population.  The birth velocity
dispersions of massive stars is of order 5 or 10 km s$^{-1}$
\citep{oh2015,aghakhanloo2017}, reflecting the gravitational potential of
the molecular clouds from which they form.  After 10 Myr,
such a velocity dispersion would mean that the star has moved 50 --
100 pc.  Fortunately, if one star disperses, many stars might disperse
as well, and in this case, the progenitor's birth companions would
still be within the central 100 pc. Even if
one considers more significant kicks, it is likely that many of the
coeval birth population received kicks, and again, the progenitor's
birth companions would be found in the same region
\citep{eldridge2011}.  Still, it is possible that the progenitor was
in a binary companion and received a large kick, propelling it
100 pc or more before collapse.

\begin{figure*}
\plotone{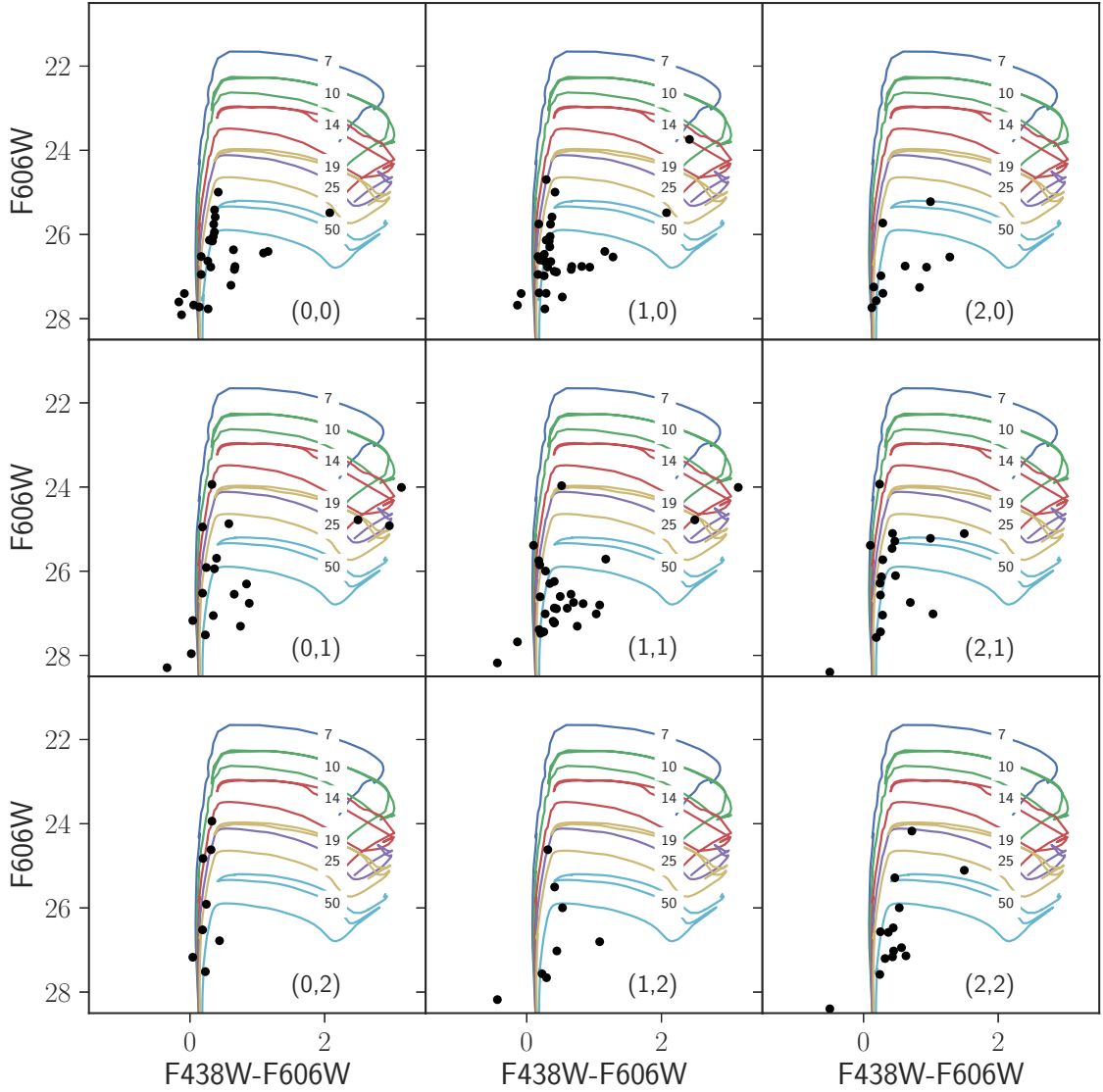}
\caption{The F438W-F606W CMD for the nine regions
  near the vanisher.  The progenitor could have been born in a
  neighboring region.  To estimate the likelihood of this, we
  calculate the SFH for the eight regions surrounding the location of
  the vanisher (central panel).  The central region has the highest number of
  bright MS stars, suggesting that it will
  have the largest and youngest SFR.}
\label{fig:CMD3x3blue}
\end{figure*}

\begin{figure*}
\plotone{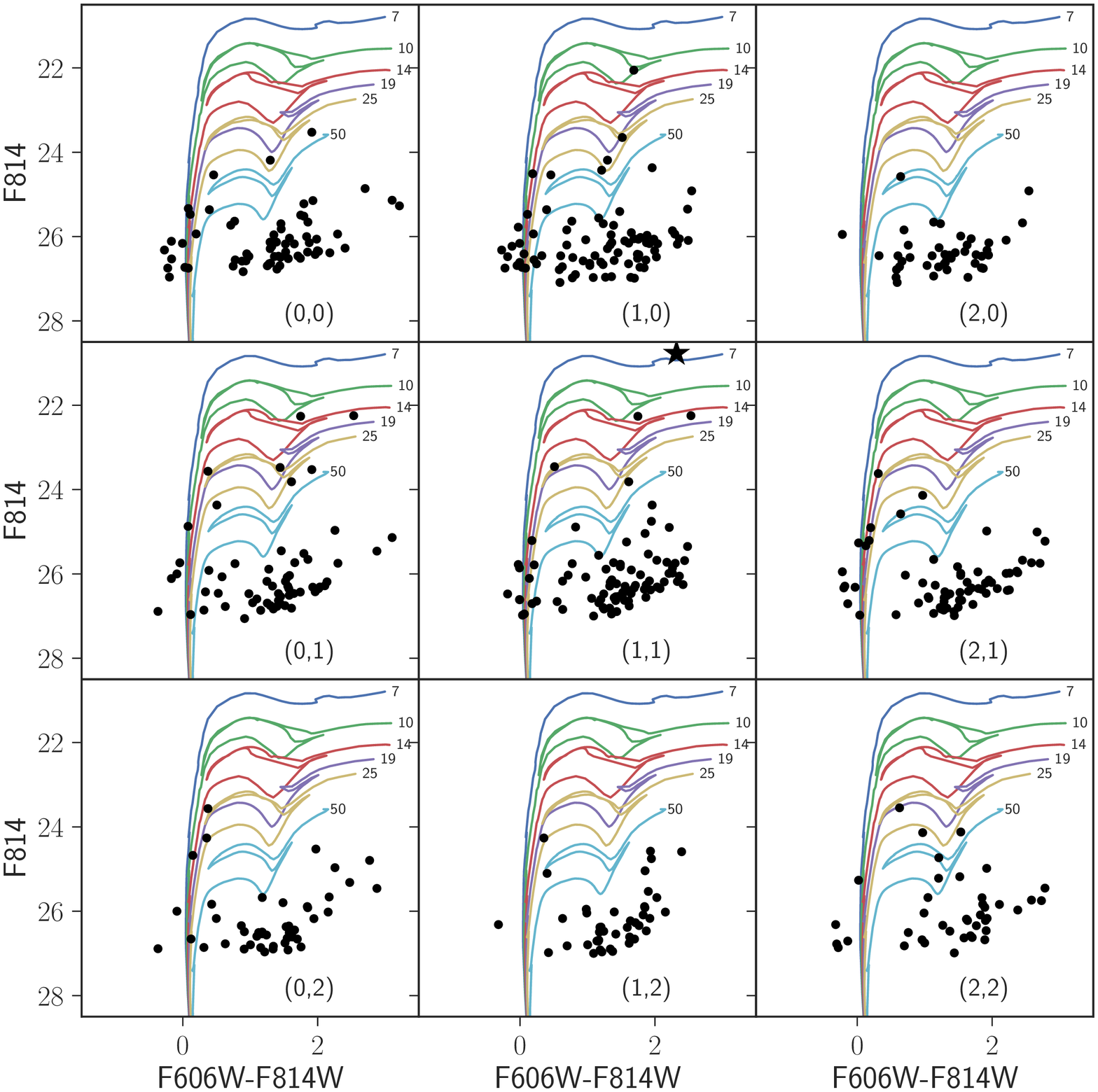}
\caption{The F606W-F814W CMD for the nine regions
  near the vanisher.  The star in the central panel is the vanisher,
  N6946-BH1.  This star is the brightest red star in all regions, a
  region that is a couple hundred pc on a side.  In fact, there
  is no star as bright as the vanisher for 300 pc.}
\label{fig:CMD3x3red}
\end{figure*}

To assess the uncertainties associated with this scenario, we calculate
the SFH and ages for the eight regions adjacent to
the central one.  Figures~\ref{fig:CMD3x3blue} and \ref{fig:CMD3x3red}
show the CMDs for the surrounding regions, along with the same
isochrones as in Figure~\ref{fig:cmds}, for reference.  Of all the CMDs, the central
region has the highest number of bright MS stars, which is a clear
indication that the central region has a best-fit SFH with the
youngest and most prominent burst of star formation.  Also, note that
the progenitor (the star in Figure~\ref{fig:CMD3x3red}) is the brightest star in all
regions, which makes it the brightest star within 100 pc.  One has to
go 300 pc before finding a star at least as bright.

\begin{figure*}
\plotone{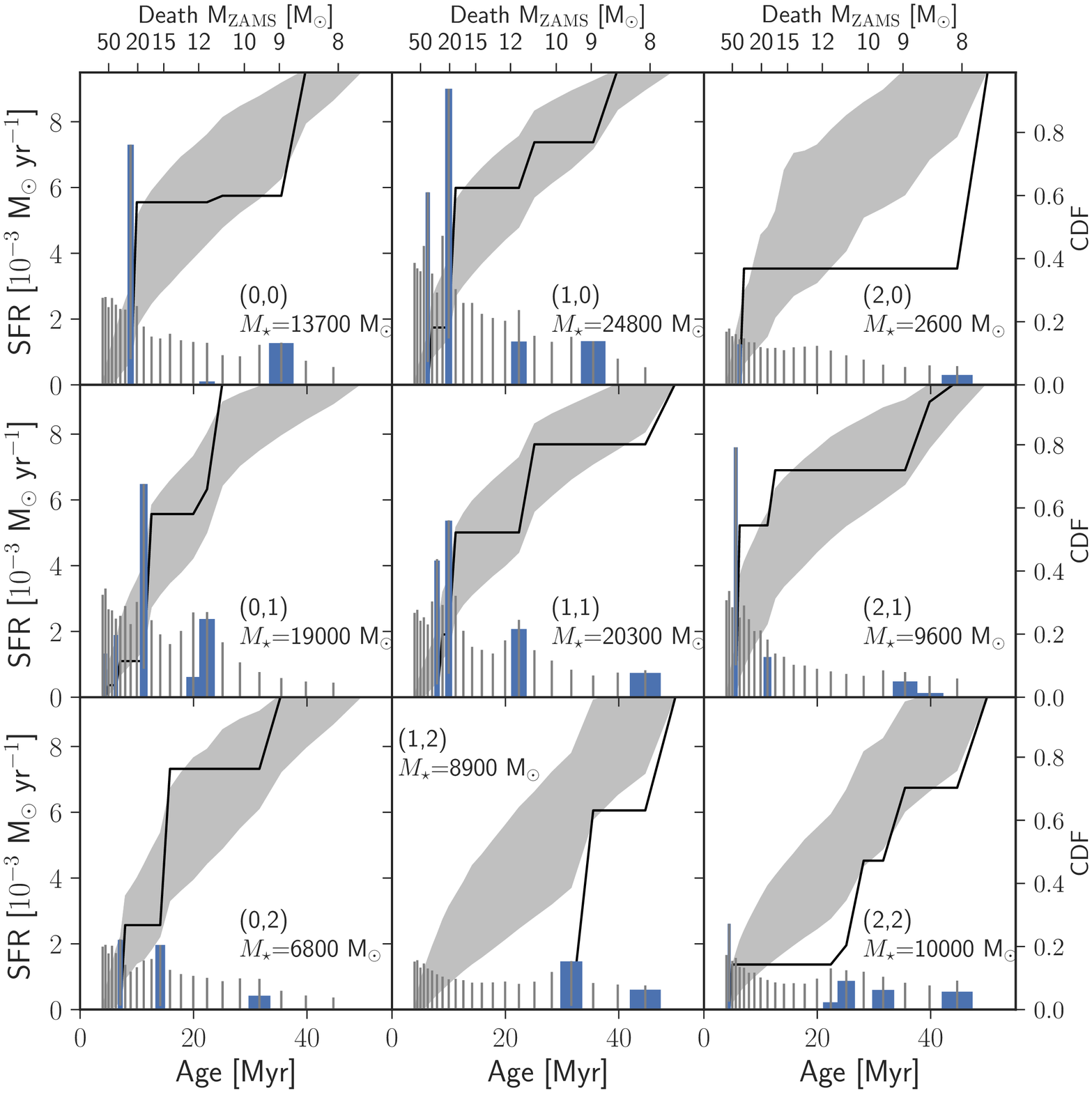}
\caption{The SFHs for the nine regions near the
  vanisher.  These plots show both the SFR (blue bars)
  and the cumulative distribution in the last 50 Myr.  Each panel
  also shows the total amount of stars formed ($M_{\star}$) in that region in the
  last 50 Myr.  The central region has the most prominent young burst of
  star formation and the most stars formed.  While there is a possibility that the vanisher was
  born in association with other stars and kicked to this region, the
  intensity of recent star formation associated with the central region suggests
  that the vanisher was born with the stars of the central region.}
\label{fig:SFH3x3}
\end{figure*}

Figure~\ref{fig:SFH3x3} shows the SFR as a function of
age for each region.  Each panel also shows the total stellar mass
formed in the last 50 Myr.  As supported by the CMDs, the most prominent burst of star formation is
the youngest burst in the central region.  Therefore, that burst's age
is also the most likely age of the progenitor.

\begin{figure*}
\plotone{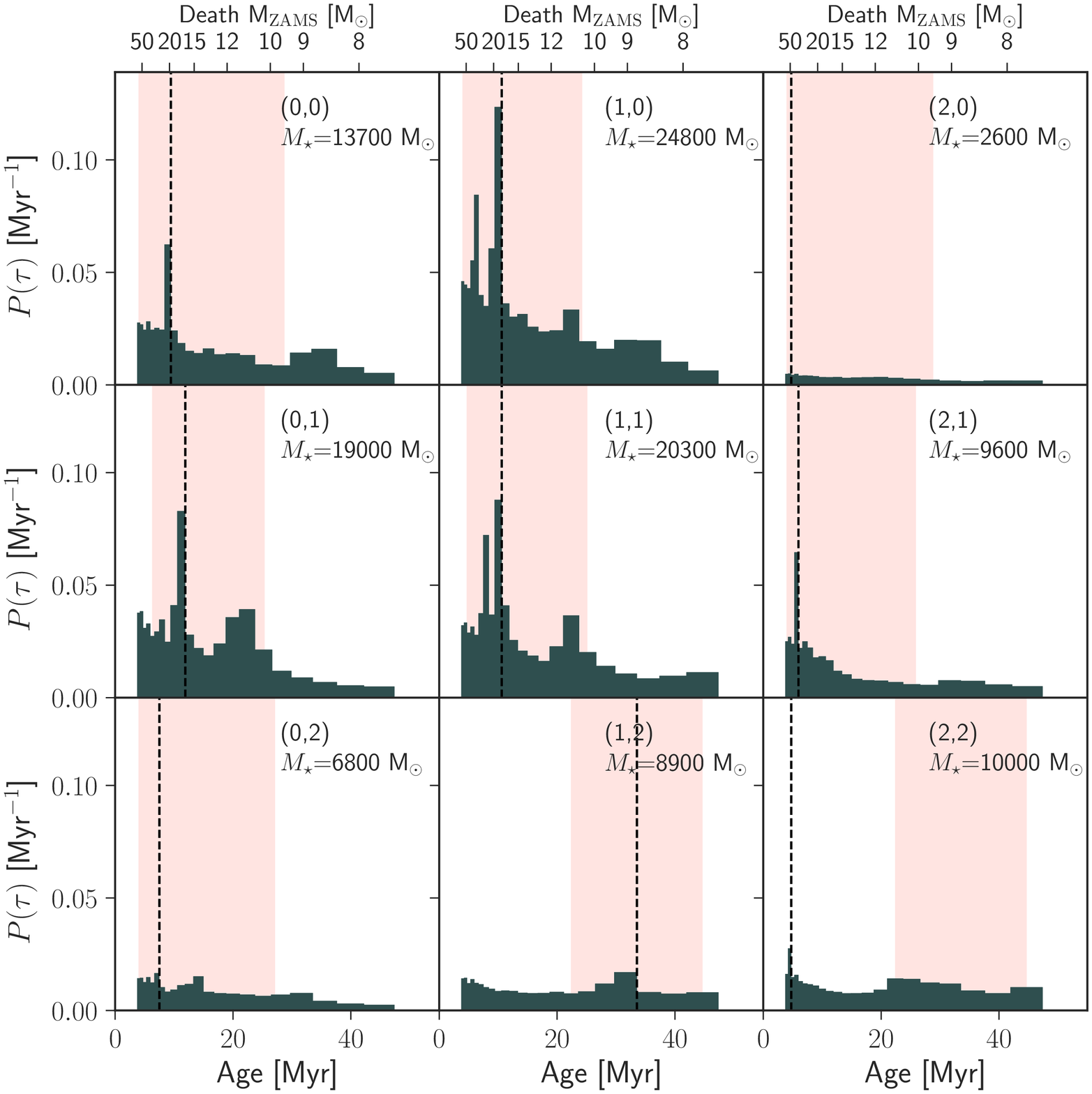}
\caption{Marginalized age distributions for the nine regions
  surrounding the vanisher.  The dashed line in each panel shows the
  most likely age, the rose vertical band highlights the narrowest
  68\% CI, and $M_{\star}$ indicates the total
  stellar mass formed in the last 50 Myr.  The central panel is
  normalized to one, and the other eight are renormalized by the
  total mass of stars formed in that region compared to the central
  region.  The central region contains the most young stars and has
  the youngest most likely age.}
\label{fig:AgePDF3x3}
\end{figure*}

Figure~\ref{fig:AgePDF3x3} shows the probability distribution for each
region's age.  The PDF in the central panel is normalized to one.
All other regions are renormalized by the ratio of their total stellar
mass ($M_{\star}$) to the central region's $M_{\star}$.  This gives a
visual representation of which regions are most likely associated with
the progenitor.  Again, the central region has the most recent and
vigorous recent SF, which reflects the fact that the central
region contains the most
number of young MS stars.  The other regions have too few stars
to really constrain their ages well.  In summary, because the
central panel has the largest number of young stars, reporting the age
and inferred mass from the central panel most likely represents the
age of the progenitor, including any possibility for dispersion or kicks.

\subsection{Comparison with SED-fitting Results}

N6946-BH1 offers a
unique opportunity to compare the SED-fitting and age-dating methods.

From SED fitting, \citet{adams2017} infer a
progenitor mass of $25^{+2}_{-1}$ $\msun$.  With the more accurate
TRGB-derived distance, the SED-derived mass becomes $27^{+2}_{-1}$
$\msun$ (the vertical black bar in Figure~\ref{fig:pdf_sfh}).  Formally, they derived an uncertainty
that is quite small, but 
this uncertainty in mass only accounts for the relatively small random
uncertainty in the photometry of the progenitor. However, the last
stages of stellar evolution are the most difficult to model, and so the
systematic uncertainties associated with modeling the SED are likely
much larger.
For example, \citet{beasor2016,beasor2018}
found that a combination of significant mass loss, dust formation,
and/or extreme bolometric correction in the last stages of red
supergiants (RSGs) can
dramatically reduce the flux in the most easily accessible visible and
near IR bands.  Hence, the
SED-to-progenitor mass mapping can be quite uncertain.
 
It is also not entirely clear that the progenitor would be in
hydrostatic equilibrium just before collapse.  The relative steadiness of the progenitor's
luminosity \citep{adams2017} supports that the progenitor was likely in a hydrostatic
state, but this is not entirely certain.  Theoretical models
suggest that there is a great deal of convective energy in the cores,
enough to unbind the outer envelope if that energy can couple
efficiently \citep{quataert2012,fuller2017}.  

The observed CMD is consistent with a
population with an age of $\bhage$ Myr.  Within the
assumption of single-star evolution, this age corresponds to an
initial mass of $\bhmass$ $\msun$.  By contrast, the SED-derived mass, $\sim$27 $\msun$,
experiences core collapse at 7 Myr \citep{marigo2017} (7.6 Myr
  for the SED-derived mass, $\sim$24, from the rotating models).  Formally,
these ages and masses are consistent within our uncertainties.  If the
progenitor is truly associated with the burst at $\sim$10 Myr, then the most likely age either
requires an $\sim$27 $\msun$ progenitor to live 40\% longer than
standard evolution theories predict or an $\sim$18 $\msun$ progenitor
must have a luminosity 2.5 times brighter than predicted just before
collapse.  
This discrepancy could be explained by either binary evolution or
episodic eruptions at the end of the progenitor's life. 

Unfortunately, in the central 100 pc, there are only a handful of
bright MS stars to constrain the age of the stellar population.
As a result, while the best-fit SFH does not show any populations
younger than 10 Myr, $\bhexclusion$ of
the age PDF (Figure~\ref{fig:pdf_sfh}) has ages older than 7 Myr, making the brightness of the progenitor marginally consistent
with the age PDF of the surrounding stars.  In this case, standard
single-star evolutionary models may be appropriate in describing the
fate of the progenitor.

\subsection{N6946-BH1 in the Context of Core-collapse Progenitor
  Theory and Observations}


In general, theory predicts that the explodability of a star depends
upon these parameters: the neutron star mass ($M_{\rm NS}$), neutron star radius
($R_{\rm NS}$), the neutrino
luminosity ($L_{\nu}$), and the amount of mass
accreting during collapse ($\dot{\mathcal{M}}$) \citep{murphy2017}.
In the context of these important parameters, there is
a critical hypersurface that divides nonexploding and explosive
solutions.  During collapse, these parameters evolve over time, and if
the evolution crosses the critical threshold, then the star explodes.
The progenitor's structure prior to collapse likely determines the
evolution in these parameters and whether the evolution crosses the
critical hypersurface for explosion.  To date, there is no direct
mapping between the progenitor structure and whether the evolution
crosses the critical hypersurface.  However, there are many studies
exploring more approximate explodability conditions \citep{burrows93,janka01,murphy08b,oconnor11,ugliano2012,ertl16,mueller16b,sukhbold2016}.

In one such study, \citet{sukhbold2016} explored the
  explodability of progenitors from 9 to 120 $\msun$; they found a general trend
  that the least massive stars explode more easily, but they also found that
  the explodability may not be entirely monotonic with mass.  More specifically,
they found islands of SN production.  Generically, their simulations
indicate that all masses below 15 $\msun$ explode.  Above this mass,
there are islands of failed explosions and BH formation.  The
existence of these islands seems to be robust to the details of stellar
evolution assumptions, but the size and frequency of these islands
depends upon progenitor prescription.  For typical stellar evolution
assumptions, the region between 15 and 22 $\msun$ shows both
explosions and failed explosions.  Between 22 and 25 $\msun$, there
seems to be a robust region of failed SNe.  Above 28 $\msun$ most
models fail to explode.  Considering all results, the minimum
progenitor to exhibit a failed SN is 15 $\msun$.  This prediction is
consistent with our best-fit progenitor mass for N6946-BH1, $\bhmass$ $\msun$.

If the explodability mostly depends upon the progenitor's initial
mass, then constraining the progenitor distribution of SNe
could constrain the the progenitor distribution of failed SNe.  For
example, \citet{smartt2015} constrained the progenitor distribution for
18 SNe IIP (plus 27 upper limits on progenitors).  Assuming a Salpeter mass function, they found a minimum
mass of $9.5^{+0.5}_{-2}$ $\msun$ and a maximum mass of $16.5^{2.5}_{-2.5}$ $\msun$.
This maximum mass for SNe IIP is consistent with the theoretical minimum mass for
failed SNe.  It is also consistent with our best-fit mass for
N6946-BH1.  However, one should note that the maximum mass for SNe IIP
may not be the maximum mass for explosion.  Progenitors above this
mass may explode as other types of SNe.  Complicating this
interpretation, \citet{davies2018} recently
reanalyzed the progenitor masses of the SNe IIP using bolometric
corrections that are derived from very red
supergiants near death.  With these new bolometric corrections, they
found a higher upper mass ($19.0^{+2.5}_{-1.3}$ $\msun$) for the
progenitor masses of SNe IIP.  

The distribution of masses from the age-dating technique does
  not show a maximum mass for explosion.  However, the distributions
  are consistent with the most massive stars failing to explode more
  often than the least massive stars.  \citet{williams2018} age-dated the stellar populations for 25
  SNe and found a distribution of median masses that is consistent
  with a wide range of masses.  However, all 25 SNe have uncertainties
  that extend below 18 $\msun$, so the masses are also consistent with
  lower masses exploding and higher masses failing to explode.  \citet{diaz-rodriguez2018} inferred
the progenitor mass distribution for 94 SNRs. They found a very
well-constrained minimum mass ($7.33^{+0.02}_{-0.16}$ $\msun$) and a power-law
slope that is steeper than Salpeter ($-2.96^{+0.45}_{-0.25}$) but a
maximum mass that is $>$59 $\msun$.  While the maximum mass is well
above the best-fit mass for N6946-BH1, the steeper-than-Salpeter
(-2.35) slope could suggest that the most massive stars explode less
frequently.

Both observations and theory of SN progenitors
  suggest that the least massive stars tend to explode more readily
  than the most massive stars.  This implies that the most massive
  stars are more likely to form black holes by a failed supernova.
  Our best-fit mass of N6946-BH1 is consistent with this general trend.


\section{Conclusion}

The vanishing star, N6946-BH1, is the first BH formation
candidate and provides the first constraints on which stars
fail to explode and form BHs.  Using stellar evolution models,
\citet{adams2017} modeled the SED and inferred an $\mzams$ of $\sim$25.
These SED-to-progenitor mass mappings rely on modeling the most
uncertain stage of stellar evolution: the last stage.  The accuracy of these models
could be dramatically affected by circumstellar dust obscuration,
extreme mass loss, or extreme bolometric corrections \citep{beasor2018,davies2018},
or even a violation of hydrostatic equilibrium \citep{quataert2012,fuller2017}.  In either case, these
uncertainties are difficult to model and quantify, making it important to constrain the progenitor mass with another technique.

We age-date the surrounding stellar population, 
which is most sensitive to modeling the MS and helium-burning phases;
both are more certain phases of stellar evolution.
We find a progenitor age of $\bhage$ Myr and an
$\mzams$ of $\bhmass$ $\msun$.  Even though the best-fit
  age-derived and the SED-derived masses differ, formally, the SED-derived mass falls within the
68\% CI of the age-derived mass.  We find that $\bhexclusion$ of our PDF lies
below this mass, hinting at a slight tension between the age and the
direct-imaging result.

To infer the age, MATCH models the magnitude and color of stars; a
fundamental parameter in modeling magnitudes is the distance.  We
use a Bayesian maximum-likelihood method to fit for the TRGB
\citep{makarov2006,mcquinn2016}.  We find a distance modulus of $m-M =
\distmodulus$, which corresponds to $\distance$ Mpc.

The primary systematic and modeling uncertainties that affect the
age-dating method are binary evolution and runaway stars.  Mergers and mass transfer modify both the mass and
stellar lifetimes in a way that is inconsistent with the predictions
of single-star stellar evolution.  If the progenitor of N6946-BH1 was a runaway star, then the age of the
nearby stars may not be coeval.  However, we derived the SFHs for
eight regions within a 100 pc, but we find that the region
colocated with the vanisher has the youngest and most active recent
star formation.  This result suggests that the vanisher is not a
runaway star.  On the other hand, the slight
difference between the most likely age and the SED-derived mass hints
that binary evolution may have played a role in the progenitor of
N6946-BH1.  But, once again, the SED-derived mass is only marginally
inconsistent with the age-derived mass.

In summary, the age of the BH formation candidate, N6946-BH1, was likely
$\bhage$ Myr.  Assuming single-star evolution, this age corresponds
to a star with a mass of $\bhmass$ $\msun$ at birth.  To better constrain the age and uncertainty will require
photometry that includes more stars in association.  This will happen
in one of three ways.  
One, a larger optical, UV, and IR space
telescope will provide more MS stars.
Two, we are lucky, and a failed SN occurs in a closer galaxy. 
Three, the region
surrounding N6946-BH1 is remarkably sparse. If the next BH formation
candidate, occurs in a more populated region, then it might be
possible to have a larger sample of main sequence stars to constrain
the age.  In any case, the next BH-formation candidate could better
constrain the progenitor and any differences between the techniques.

\section*{ACKNOWLEDGMENTS}
Based on observations made with the NASA/ESA Hubble Space Telescope,
obtained from the Data Archive at the Space Telescope Science
Institute, which is operated by the Association of Universities for
Research in Astronomy, Inc., under NASA contract NAS 5-26555. These
observations are associated with programs GO-14266 and
GO-13392. Support for programs HST-AR-15042 and HST-GO-14786 was
provided by NASA through a grant from the Space Telescope Science
Institute, which is operated by the Association of Universities for
Research in Astronomy, Inc., under NASA contract NAS 5-26555.


\end{document}